\documentclass[amssymb,amsmath,letterpaper]{article}

\usepackage{opex3}
\usepackage{graphicx}
\usepackage{dcolumn}
\usepackage{bm}
\usepackage{amsmath}

\begin{document}

\title{Strong magneto-optical effects due to surface states in three-dimensional topological insulators}

\author{Xianghan Yao$^1$, Mikhail Tokman$^2$, and Alexey Belyanin$^{1,*}$}

\address{$^1$Department of Physics and Astronomy, Texas A\&M 
University, College Station, TX 77843 USA \\ $^2$ Institute of Applied Physics, Russian Academy of Sciences, Nizhny Novgorod 603950, Russia}

\email{$^*$belyanin@tamu.edu}


\begin{abstract}
We show that a thin film of a three-dimensional topological insulator such as Bi$_2$Se$_3$ or Bi$_2$Te$_3$ can exhibit strong linear and nonlinear magneto-optical effects in a transverse magnetic field. In particular, one can achieve an almost complete circular polarization of an incident terahertz or mid-infrared radiation and an efficient four-wave mixing.
\end{abstract}


\section{Introduction}

Three-dimensional topological insulators (TIs) such as Bi$_2$Se$_3$ or Bi$_2$Te$_3$ are ordinary gapped insulators in the bulk with gapless chiral Dirac fermion states on the surface that show a number of fascinating properties \cite{review,zhang} and promise applications in electronic devices as well as terahertz and infrared optics \cite{THz}. The Fermi velocity for low-energy excitations near the Dirac point is close to that in graphene, which gives rise to similar matrix elements of the optical transitions between surface states. Unlike the Dirac fermions in monolayer graphene, surface states are not spin-degenerate. The spin direction is locked to the direction of momentum forming a chiral structure. Momentum scattering of carriers in surface states requires a spin flip and is suppressed due to the time-reversal symmetry. This leads to a long scattering time in high-quality samples without magnetic impurities, which enhances both linear and nonlinear optical response.  

Applying a magnetic field breaks the time-reversal symmetry but the momentum scattering is still partially suppressed. Recent experiments in a strong transverse magnetic field have revealed the Landau levels of massless Dirac fermions states with energies $E_n$ that scale as $\sqrt{B|n|}$, and demonstrated their longer decoherence time as compared to the Dirac fermion states in graphene \cite{LLs1,LLs2,LLs3}. This leads to sharper transition lines and stronger light-matter coupling, giving rise to strong magneto-optical effects such as a giant Kerr rotation $\theta_K \sim 1$ rad at THz wavelengths \cite{aguilar}.  Furthermore, some of these materials have a relatively large bulk band gap (0.3 eV for 
Bi$_2$Se$_3$ and 0.2 eV for Bi$_2$Te$_3$ \cite{zhang,hsieh}) and a tunable Fermi level which can be put within the bulk gap close to the Dirac point. Therefore, long-wavelength mid-infrared and THz radiation is not affected by interband absorption between the bulk states and can be selectively coupled to surface states. 

In this paper we concentrate on the polarization and nonlinear optical effects in TI films. We show that high-quality thin films can provide a nearly complete circular polarization of the incoming radiation and lead to strong four-wave mixing effects in the mid-infrared and THz range. Their thickness is constrained from below by electron tunneling which leads to the opening of a band gap. This means that the films can be as thin as several nanometers. In Section II we describe the electron states and linear optical properties of the surface states in Bi$_2$Se$_3$, both with and without magnetic field. We calculate the polarization coefficient for TI films showing the possibility of a near-complete circular polarization of an incoming radiation. In Section III we present the results on the four-wave mixing and stimulated Raman scattering of TI films. 

 \section{Linear optical properties of the surface states}

We start with the effective Hamiltonian for the topological insulator Bi$_2$Se$_3$ without a magnetic field. It has been considered a number of times; here we give a brief summary of the main results relevant for the optical properties. We will use the parameters from \cite{parameters}. Following the approach in Ref.~\cite{thinfilm_qian,thinfilm_shen}, we use the hybridized states of Se and Bi orbitals $\{|p1^{+}_{z},\uparrow\rangle, |p2^{-}_{z},\uparrow\rangle, |p1^{+}_{z},\downarrow\rangle, |p2^{-}_{z},\downarrow\rangle\}$. The Hamiltonian is given by
\begin{equation}
H(k)=(C-D_1\partial^2_z+D_2k^2)+\left(\begin{array}{cc}
h(A_1) & A_2k_-\sigma_x \\
A_2k_+\sigma_x & h(-A_1)
\end{array}\right)
\end{equation}
where
$$
h(A_1) = (M + B_1\partial^2_z - B_2k^2)\sigma_z - iA_1\partial_z\sigma_x,
$$
$k_{\pm} = k_x \pm ik_y$ and $k^2 = k_x^2 + k_y^2$. The parameters are \cite{parameters}: $M =0.28$ eV, $A_1=2.2$   eV $\AA$, $A_2=4.1$ eV $\AA$, $B_1=10$ eV $\AA^2$, $B_2=56.6$ eV $\AA^2$, $C=-0.0068$ eV, $D_1=1.3$  eV $\AA^2$, $D_2=19.6$ eV $\AA^2$. By expanding the Hamiltonian with the solutions of the surface states at the $\Gamma$ point, the Hamiltonian can be expressed in a block-diagonal form
\begin{equation}
H(k) = \left(\begin{array}{cc}
h_+(k) & 0 \\
0 & h_-(k)\end{array}\right);
\end{equation}
with 
$$
h_\pm(k) = E_0 - Dk^2 + \hbar\upsilon_F\left( \vec{\sigma} \times \vec{k} \right)_z \pm \sigma_z\left(\Delta/2 - Bk^2\right),
$$
where $E_0 = (E_+ + E_-)/2, \Delta = E_+ - E_-$. Here $E_+$ and $E_-$ are the energies of surface states at the Dirac point. The explicit expressions for $E_0, D, \Delta$ and $B$ are thickness-dependent and can be found in Ref.~\cite{thinfilm_qian,thinfilm_shen}. 
When the film thickness is greater than about 6 quintuple layers (QLs), i.e. greater than 6 nm, the electron coupling between top and bottom surface states becomes weak and the above parameters become constant with $E_0 = 0.0337$ eV, $D = -12.25$ eV$\AA^2$, $\hbar\upsilon_F = 4.07$ eV $\AA$, and $\Delta$ and $B$ almost zero. As a result, the surface states becomes gapless in the vicinity of the Dirac point, and $h_+(k) = h_-(k)$. We then use only one block as the effective Hamiltonian
\begin{equation}\label{Hss}
H_{eff}(k) = E_0 - Dk^2 + \hbar\upsilon_F(k_y\sigma_x - k_x\sigma_y),
\end{equation}
and add a surface degeneracy factor of $g_s = 2$ when calculating the optical response for the radiation interacting with both surfaces. The energy bands are shown in Fig.~1. It is clear that for low-energy excitations up to $\sim 100$ meV the $k^2$ term can be neglected. 

If a TI film is placed in a uniform perpendicular magnetic field, its effect on the orbital motion is included by the Perierls substitution $\vec{\pi} = \vec{k} + \frac{e}{\hbar c}\vec{A}$. The new Hamiltonian is introduced using annihilation and creation operators $a = \frac{l_c}{\sqrt{2}}\pi_-$ and $a^\dag = \frac{l_c}{\sqrt{2}}\pi_+$. Here $l_c = \displaystyle \sqrt{\frac{\hbar c}{eB}}$ is the magnetic length and $\pi_{\pm} = \pi_x \pm i \pi_y$.
\begin{equation}
\label{eff} 
H_{eff}(\pi) = E_0 - \frac{2D}{l^2_c}(a^\dag a + \frac{1}{2}) + \frac{\sqrt{2}\hbar\upsilon_F}{l_c}\left(\begin{array}{cc}
0 & ia \\
-ia^\dag & 0\end{array}\right).
\end{equation}
The eigenfunctions and eigenvalues can be solved together with $a^\dag\psi_{|n|} = \sqrt{|n|+1}\psi_{|n|+1}$,  $a\psi_{|n|} = \sqrt{|n|}\psi_{|n|-1}$.
\begin{eqnarray}
\label{state}
&&\Psi_{n} = \frac{C_n}{\sqrt{L}}e^{-ik_yy} \left(
{\rm sgn}(n)\cdot{\rm i}^{|n|}\psi_{|n|-1},~
 {\rm i}^{|n|-1}\psi_{|n|}\right)^T,\nonumber\\
&&E_n = E_0 - \frac{2|n|D}{l_c^2} + {\rm sgn}(n)\sqrt{2|n|(\frac{\hbar\upsilon_F}{l_c})^2+(\frac{D}{l_c})^2}\nonumber.\\
\end{eqnarray}
Here $C_0 = 1/\sqrt{2}$, $C_{n\ne0} = 1$, and $\psi_{|n|}$ is orthogonal Hermite polynomials. 
The magnetic field condenses the continuous k-dependent states into discrete surface Landau states, and brings a k-degeneracy of $\frac{1}{2\pi l_c^2}$. 

For low-energy excitations in the terahertz or long-wavelength mid-infrared region below 100 meV, it is safe to only keep linear terms in the Hamiltonian as the second-order correction is small; see Fig.~1(a). When the terms with $D$ are neglected, the eigenenergies in Eq.~(\ref{state}) have the same universal behavior as in graphene. 
Therefore, in the following discussion  we will use a simplified linear Hamiltonian
\begin{equation}
H_{eff}(\pi) = E_0 - \hbar\upsilon_F(\pi_x\sigma_y - \pi_y\sigma_x).
\end{equation}

\begin{figure}[htb]
\centerline{
\includegraphics[width=9cm]{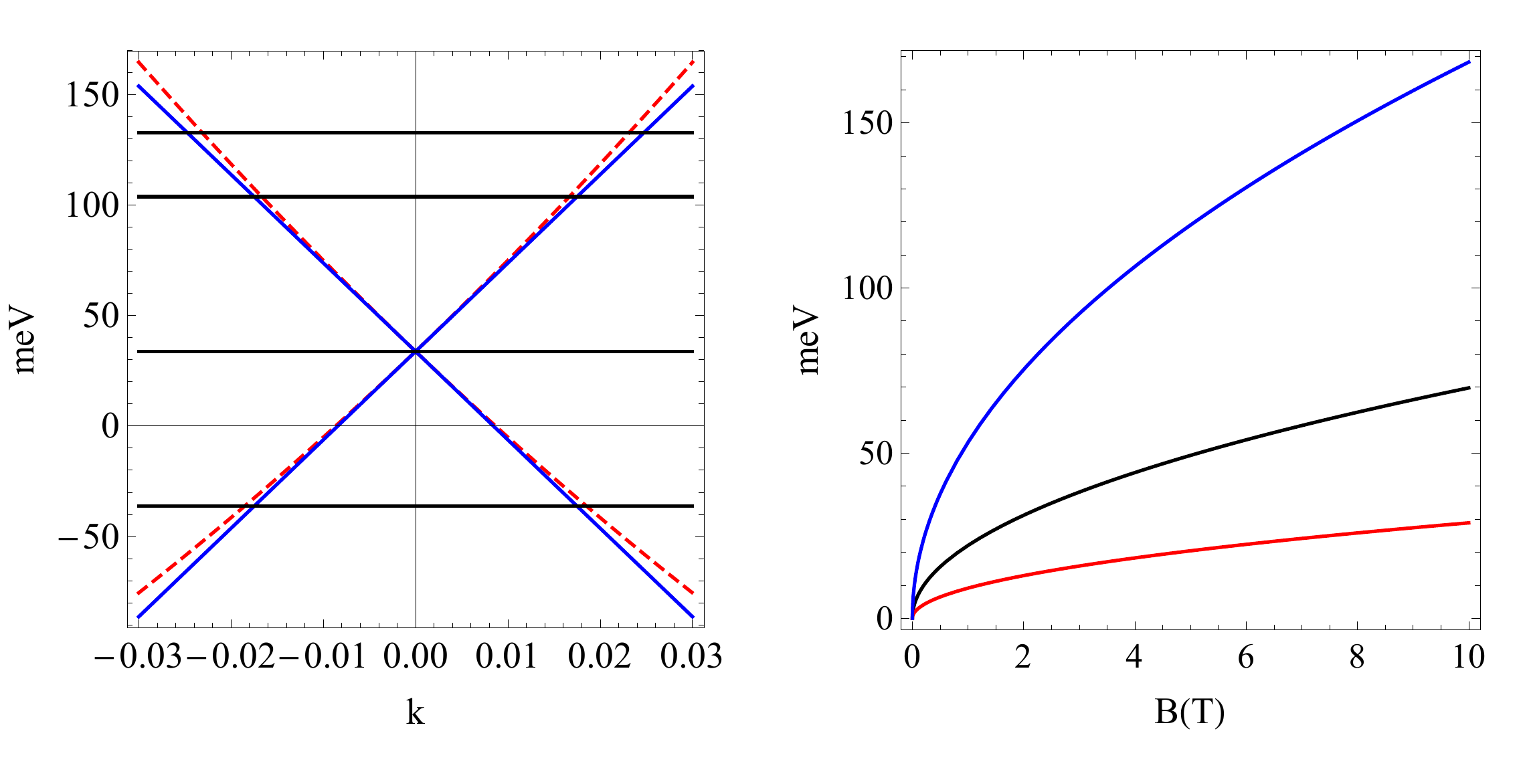}}
\caption{(a) Energy bands of the surface states in Bi$_2$Se$_3$ near the $\Gamma$ point at zero magnetic field (with quadratic correction shown by dashed line), and energies of Landau levels in a magnetic field of 10 T (horizontal lines).
(b) Landau level energies as a function of the magnetic field for Bi$_2$Se$_3$. }
\end{figure}

After finding energies and wave functions for the surface states of topological insulators in a strong magnetic field one can calculate the optical response.  
For a normally incident optical field, the interaction Hamiltonian describing the coupling between the surface Landau states and light is
\begin{equation}
H_{int} = \upsilon_F(\vec{\sigma}\times\vec{A}_{opt})_z \equiv -(\tilde{\vec{\mu}}\times\vec{E}(\omega))_ze^{-i\omega t},
\end{equation}
where $\tilde{\vec{\mu}}$ is defined as $\frac{ie\upsilon_F}{\omega}\vec{\sigma}$. 
The  matrix element of the dipole moment corresponding to the optical transition between states $n$ and $m$ can be calculated as
\begin{equation}
\vec{\mu}_{mn} = e\langle m |\vec{r}|n\rangle = \frac{ie\hbar\upsilon_F}{E_n-E_m}\langle m |\sigma_x\hat{y} - \sigma_y\hat{x}|n\rangle.
\end{equation}
It has the same magnitude and linear scaling with  $\lambda$ as the one in monolayer graphene. The selection rules are similar to that for graphene. Namely, $\Delta |n| = \pm 1$, and $\hat{e}_{RHS}$ photons are absorbed when $|n_f| = |n_i| - 1$, while $\hat{e}_{LHS}$ photons are absorbed when $|n_f| = |n_i| + 1$. The transition frequencies between  LLs are in the mid/far-infrared region when the magnetic field strength is of the order of a few Tesla, as shown in Fig. 1(b).

High-frequency absorbance in a TI film in the absence of a magnetic field has a constant value $\frac{\pi}{2}\cdot\frac{e^2}{\hbar c}$ \cite{THz}, where we included a degeneracy factor of 2 stemming from two surfaces. In a quantizing magnetic field transitions between the Landau levels (LLs) give rise to multiple cyclotron resonances with peak absorbance values scaling as $\displaystyle \frac{\omega}{\gamma}\cdot\frac{e^2}{\hbar c}$,
where $\gamma$ is a line broadening. 

 Here we use the density matrix formalism to calculate the linear susceptibility and magneto-optical effects due to the surface states in a TI film.
 The 2D optical polarization is defined as
 \begin{equation}\label{P}
\vec{P}(\vec{r},t) = N\cdot{\rm tr}(\hat{\rho}\cdot\vec{\mu}).
\end{equation}
where $N$ is the surface density of states per each surface LL. The density matrix elements $\rho_{nm}$ are calculated from the master equation with phenomenological relaxation rates $\gamma_{nm}$:
\begin{eqnarray} \label{rho}
 \dot{\rho}_{nm}&=&-\frac{i}{\hbar}(\varepsilon_n - \varepsilon_m)\rho_{nm}-\frac{i}{\hbar}[ \hat{H}_{int}(t), \hat{\rho} ]_{nm}
 \nonumber \\
 &-&\gamma_{nm}(\rho_{nm}-\rho_{nm}^{(eq)}).
\end{eqnarray}

For a given incident field $\vec{E}(\omega) = E(\omega)e^{-i\omega t}\hat{e}$, using the 1st-order perturbation solution for a density matrix, the corresponding resonant part of the polarization per one surface layer of a TI is
\begin{eqnarray}
\label{linearP}
\vec{P}^{(1)}(\omega)
&=& N \sum_{nm} \frac{\rho_{mm}^{(eq)}-\rho_{nn}^{(eq)}}{\hbar}\cdot\frac{\left(\tilde{\vec{\mu}}_{nm}\times\hat{e}\right)_z\vec{\mu}_{mn}}{(\omega_{nm}-\omega)-i\gamma_{nm}}\nonumber\\
&\cdot&E(\omega)\exp{(-i\omega t)} .
\end{eqnarray}

\subsection{Polarization effects for a thin film}

If the thickness of a TI film is much smaller than the wavelength of incidence {\it and} in the limit $\frac{\omega_c}{\gamma}\frac{e^2}{\hbar c}\ll1$,
one can directly apply standard formulas for a weak absorption and Faraday rotation of a linearly polarized light \cite{LL8}. 
When $\frac{\omega_c}{\gamma}\frac{e^2}{\hbar c}>1$, a TI film becomes optically thick in the centers of the cyclotron lines. We should no longer use the standard formulas, but instead solve Maxwell's equations together with an induced surface current
$\vec{j}_{\perp} = -i\omega\tilde{\chi}_{\omega}\vec{E}_{\perp}$. Here $\vec{E}_{\perp}$ is the in-plane electric field and $\tilde{\chi}_{\omega}$ is the surface (2D) susceptibility tensor. Consider an incident field to be close to resonance  with one particular  transition between states $n$ and $m$ ($\omega \approx \omega_{nm}$). Then the linear optical response is dominated by this particular transition:
\begin{equation}
\label{chi_ij}
\chi^{(1)}_{kj}(\omega) = N\frac{\rho_{mm}^{(eq)}-\rho_{nn}^{(eq)}}{2\hbar}
\frac{\epsilon_{zi'j}\tilde{\mu}^{i'}_{nm}\mu^k_{mn}}{\omega_{nm}-\omega-i\gamma},
\end{equation}
where $\epsilon_{ijk}$ is the Levi-Civita symbol and there is a summation with respect to index $i'$.  
In the geometry of Fig. 2(a) the resulting linear susceptibility tensor is in a gyrotropic form:
\begin{equation}
\tilde{\chi} =\left(
\begin{array}{cc}
\chi_{xx}& \chi_{xy}\\
\chi_{yx}&\chi_{yy}\end{array}\right)
=
\left(
\begin{array}{cc}
\chi_{\perp}& -ig\\
ig&\chi_{\perp}\end{array}\right),
\end{equation}
where $\chi_{\perp}$ and $g$ can  be calculated from Eq.~(\ref{chi_ij}) as
\begin{eqnarray}
\chi_{\perp} &=& \frac{\Omega^2_{nm}}{\omega^2_{nm}-(\omega+i\gamma)^2};\nonumber\\
g &=& s\cdot \frac{\omega+i\gamma}{\omega_{nm}}\frac{\Omega^2_{nm}}{\omega^2_{nm}-(\omega+i\gamma)^2},
\end{eqnarray}
with $$\Omega^2_{nm} = \frac{C^2_mC^2_n(\rho_{mm}^{(eq)}-\rho_{nn}^{(eq)})Ne^2\upsilon^2_F}{\hbar\omega},$$
$$
s = \left\{\begin{array}{cc}
+1, & |n|=|m|-1\\
-1, & |m|=|n|-1\end{array}\right. ; \quad C_n=\left\{\begin{array}{cc}
1 &(n=0) \\
\frac{1}{\sqrt{2}} &(n\ne0)
\end{array}\right. .
$$

\begin{figure}[htb]\label{K_a}
\centerline{
\includegraphics[width=12cm]{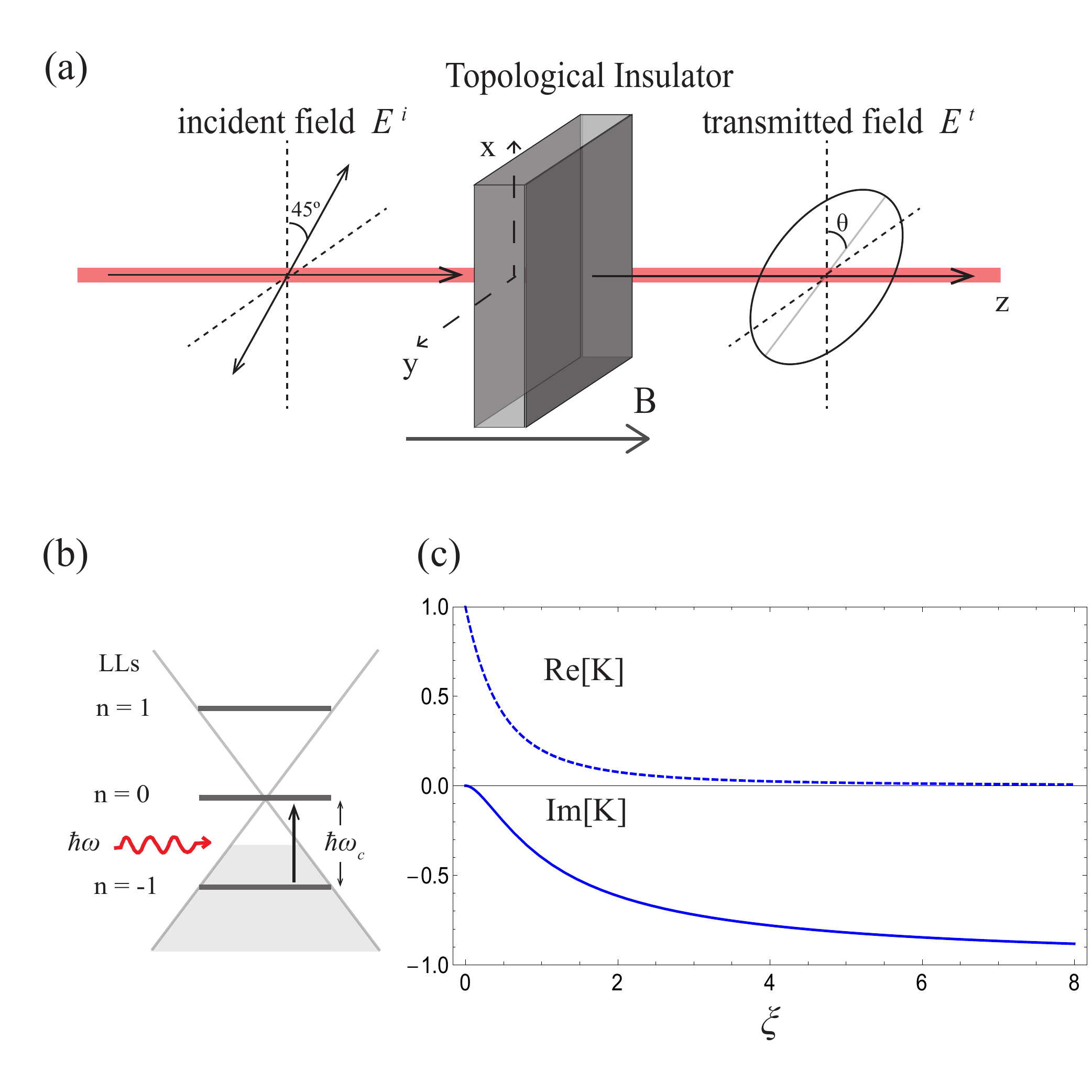}}
\caption{(a) An example of the experimental geometry: the incident field is linearly polarized with orientation angle $\frac{\pi}{4}$. (b)The optical transition scheme for incident frequency $\omega \approx \omega_c$. Here Fermi level is placed between Landau levels -1 and 0. (c) Polarization coefficient $K$ of the transmitted optical field, as a function of $\xi = \frac{\omega_c}{4\gamma}\frac{e^2}{\hbar c}$. The slab thickness is chosen as 0.01 $\lambda$ in the plot.}
\end{figure}

For the radiation normally incident on a TI film or graphene on a substrate (see Fig.~2(a)), with an in-plane electric field given by  $\tilde{E}_{\perp}=\left(E_y, E_x\right)^T$, the transmitted field is:
\begin{equation}
\left(\begin{array}{c}
E'_y\\
E'_x\end{array}\right)
=\frac{
\left(\begin{array}{cc}
1-\alpha\chi_{xx}&\alpha\chi_{yx}\\
\alpha\chi_{xy}  &1-\alpha\chi_{yy}\end{array}
\right)}{(1-\alpha\chi_{xx})^2+(\alpha\chi_{xy})^2}
\left(\begin{array}{c}
E_y\\
E_x\end{array}\right),
\end{equation}
where $\alpha\equiv i\frac{2\pi\omega}{c}$. If we define the in-plane polarization coefficient as $K\equiv E_y/E_x$, the corresponding polarization coefficient for the transmitted wave is
\begin{equation}
\label{ellipse}
K'=\frac{K(1-\alpha\chi_{xx})+\alpha\chi_{yx}}{K\alpha\chi_{xy}+(1-\alpha\chi_{yy})}.
\end{equation}
The real and imaginary parts of $K'$ are shown in Fig.~3(c). As shown in the figure (see also Fig.~4(a)), in the limit $\frac{\omega_c}{\gamma}\frac{e^2}{\hbar c}>1$ $K \rightarrow -i$, i.e. the field component with resonant polarization will be almost completely reflected and only the non-resonant circular polarization will go through, thus resulting in a nearly complete circular polarization of the transmitted radiation. Although Figs.~2(c) and 4(a) are plotted for the film thickness $d = 0.01 \lambda$, the same result is obtained for smaller film thicknesses as long as $d > 6$ nm. 

Although we considered only normal incidence, for a thin layer Eq.~(\ref{ellipse}) remains valid for an obliquely incident light. In this case it describes the projection of the electric field vector on the plane of the layer. By varying the angle of incidence $\alpha$ one can change the eccentricity of the polarization ellipse. For example, if the transmitted polarization at normal incidence is a circle, the ratio of axes of an ellipse scales as $a/b = \cos\alpha$ with increasing $\alpha$.  

\subsection{Polarization effects in a slab geometry}

A slab geometry with two 2D massless fermion layers on opposite surfaces emerges naturally for a thick TI film and can also be implemented by putting graphene layers or two thin TI films on two sides of a dielectric substrate; see Fig.~3(a).  For a slab thickness $d$  comparable to the wavelength one has to take into account multiple optical reflections between the two surface layers. 
For an optical field normally incident from $z<0$ onto the slab, one can similarly derive the transmission matrices for the lower and upper surface:

\begin{figure}[htb]\label{K_L}
\centerline{
\includegraphics[width=12cm]{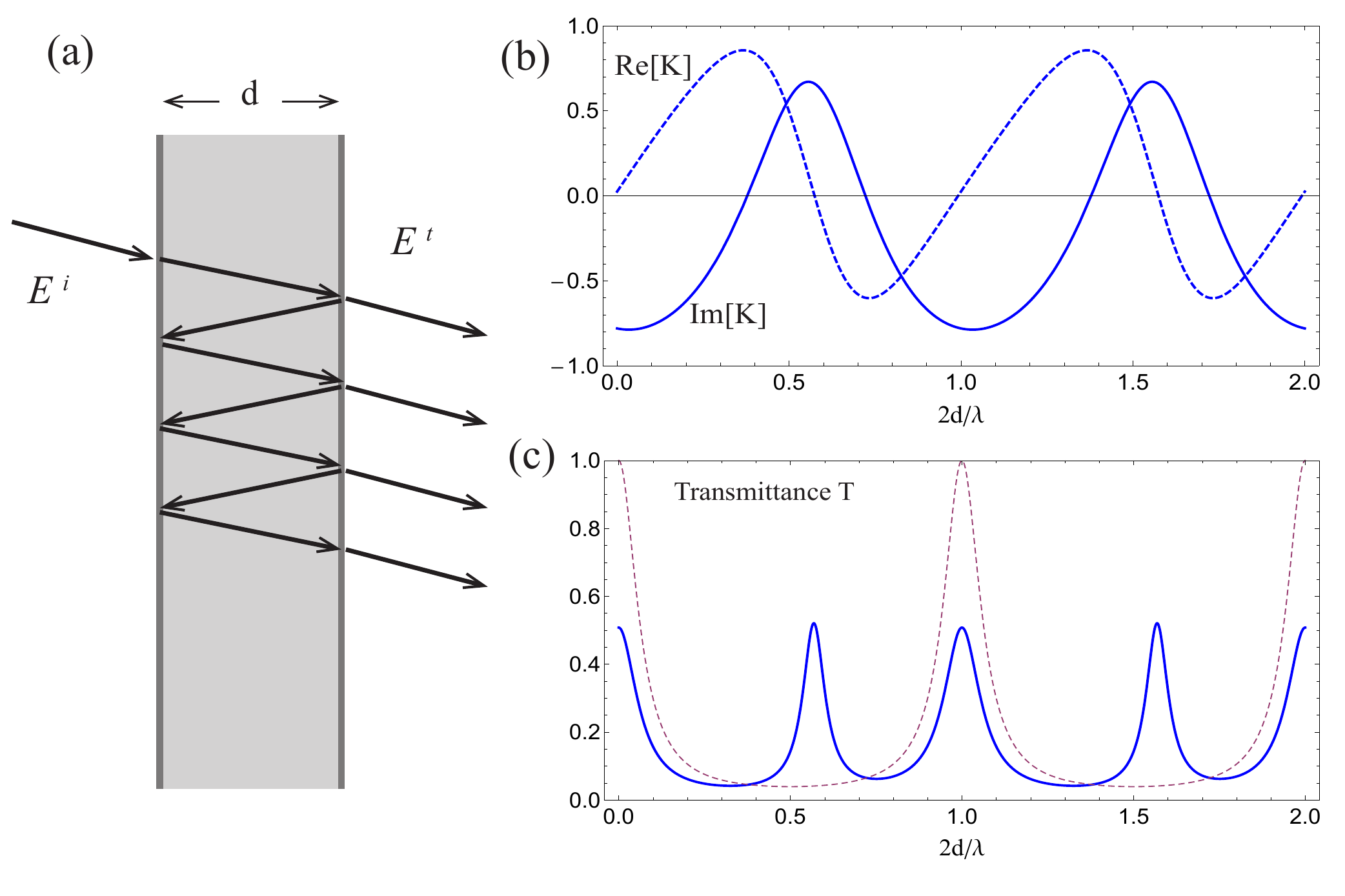}}
\caption{(a) Multiple reflections in the slab geometry. (b) The real and imaginary part of the polarization coefficient (K) of the transmitted optical field as a function of the slab thickness $d$. (c) Transmittance T as a function of slab thickness $d$ (solid line). The dashed line is for a pure dielectric slab without surface layers of massless fermions. Here $\xi = 4$ in both plots.}
\end{figure}

\begin{eqnarray}
T_1&=&\frac{2
\left(\begin{array}{cc}
n+1-2\alpha\chi_{\perp} & 2\alpha ig\\
-2\alpha ig & n+1-2\alpha\chi_{\perp}\end{array}\right)}
{(n+1-2\alpha\chi_{\perp})^2-(2\alpha g)^2}\\
T_2&=&\frac{2
\left(\begin{array}{cc}
n^2+n-2n\alpha\chi_{\perp} & 2n\alpha ig\\
-2n\alpha ig & n^2+n-2n\alpha\chi_{\perp}\end{array}\right)}
{(1+n-2\alpha\chi_{\perp})^2-(2\alpha g)^2}
\end{eqnarray}
where $n$ is a high-frequency refractive index of a bulk material between the surfaces. The transmitted in-plane electric field is
\begin{equation}
\left(\begin{array}{c}
E'_y\\
E'_x\end{array}\right)
=e^{\frac{in\omega L}{c}}T_2\left(I-e^{2\frac{in\omega L}{c}}R_2R_2\right)^{-1}T_1
\left(\begin{array}{c}
E_y\\
E_x\end{array}\right)
\end{equation} 
As can be seen in Fig.~3(b,c), the standard Fabry-Perot transmission peaks at $2d/\lambda = N$ where $N = 1,2, ...$ correspond to the points of near-complete circular polarization when $K \rightarrow -i$, similarly to the case of a thin film $d \ll \lambda$. In addition, surface layers give rise to extra peaks at $2d/\lambda \approx 1/2, 3/2, ...$ where the transmitted field has a circular polarization with an opposite sense. Therefore, the same material can produce a beam circularly polarized in both directions by changing $d$ or resonant wavelength $\lambda$. This is illustrated in Fig.~4(b). 

\begin{figure}[htb]
\centerline{
\includegraphics[width=16cm]{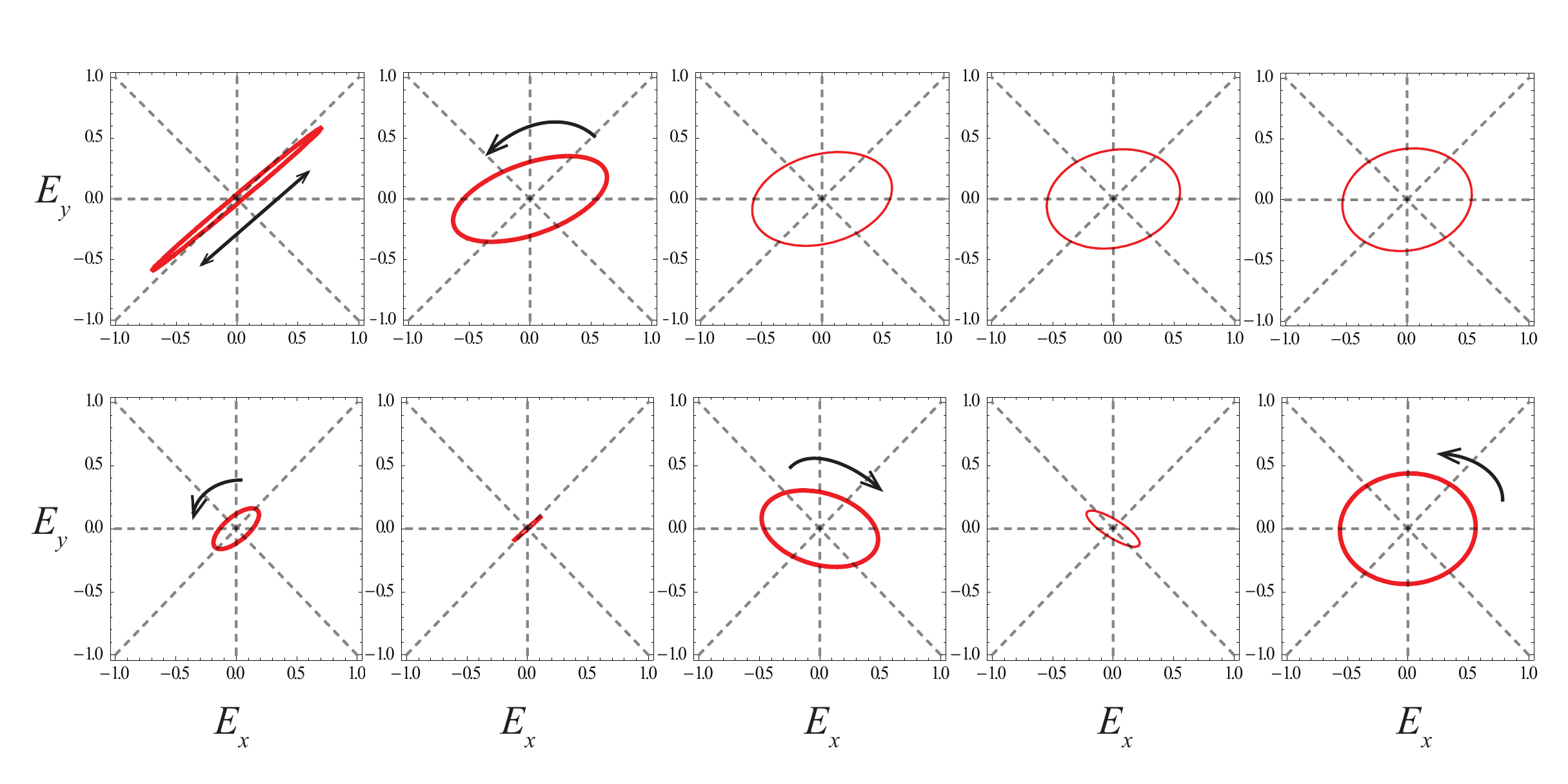}}
\caption{Polarization ellipse of the transmitted field. First row: from left to right, $\xi = 0.1, 1, 2, 3, 4$; $d$ is fixed at $0.01\lambda$. Second row: from left to right, $d = 0.01\lambda,  0.2\lambda,  0.29\lambda,  0.4\lambda,  0.5\lambda$; $\xi$ is fixed at 4. The sense of rotation in the polarization is indicated by arrows. }
\end{figure}

\section{Nonlinear optical properties}

Although there are clear differences between 2D electron states in the  topological insulators and in graphene that one can see e.g. in the chiral structure of the Hamiltonian (\ref{eff}) and its eigenenergies/eigenstates (\ref{state}), they share the same $\pm \sqrt{|n|B}$ sequence  of the LLs and the same structure of the dipole matrix elements. This results in similar nonlinear optical properties. Here we consider just one example: a resonant four-wave mixing process based on resonant transitions between the LLs of surface states in Bi$_2$Se$_3$ (Fig.~5) which is similar to the one considered in \cite{yao} and shows similar nonlinear conversion efficiency. Another four-wave mixing process of an efficient two-photon parametric decay in a four-level scheme of LLs was  considered in \cite{tokman}.   

\begin{figure}[htb]
\centerline{
\includegraphics[width=14cm]{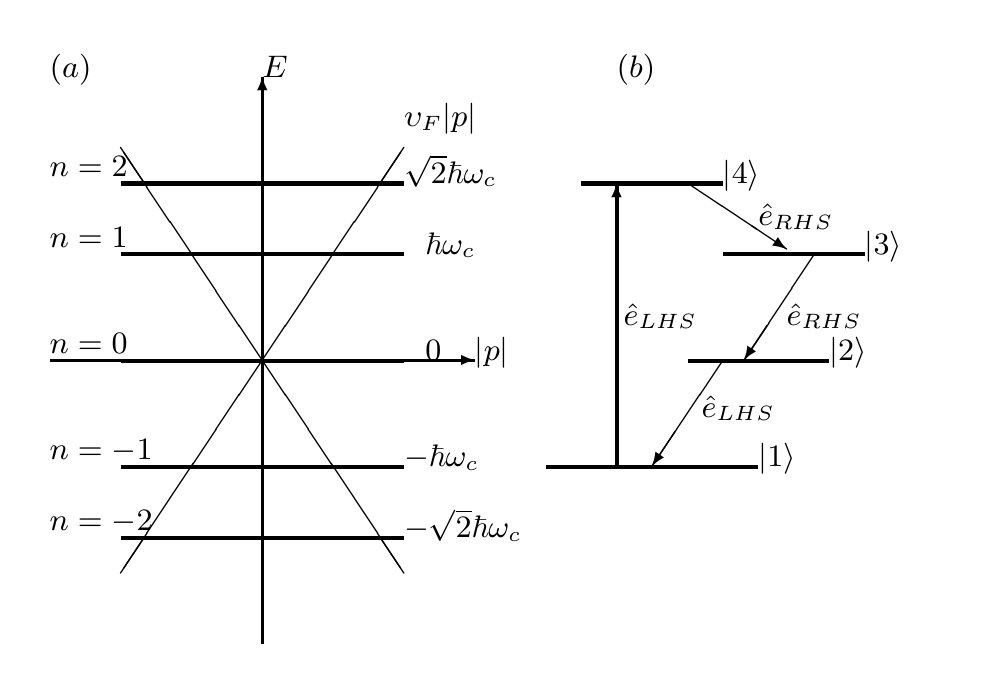}}
 \caption{Landau levels near the Dirac point superimposed on the electron dispersion without the magnetic field $E=\pm\upsilon_F |p|$. (b): A scheme of the four-wave mixing process in the four-level system of Landau levels with energy quantum numbers $n = -1, 0, +1, +2$ that are renamed to states 1 through 4 for convenience. }\label{4wscheme}
\end{figure}

In Fig.~5 all transitions connected by arrows are allowed and the dipole moment matrix of the 4-level-system is given by 
\begin{equation}
\vec{\mu}= \displaystyle  \frac{el_c}{2}
\left(
\begin{array}{cccc}
 0&-i\hat{x}-\hat{y}&0&\frac{i\hat{x}-\hat{y}}{2+\sqrt{2}}  \\
 i\hat{x}-\hat{y}&0&i\hat{x}-\hat{y}&0 \\
 0&-i\hat{x}-\hat{y}&0&\frac{i\hat{x}-\hat{y}}{2-\sqrt{2}}  \\
 \frac{-i\hat{x}-\hat{y}}{2+\sqrt{2}}&0&\frac{-i\hat{x}-\hat{y}}{2-\sqrt{2}}&0
\end{array}\right)
\end{equation}

Consider a strong bichromatic normally incident field $\vec{E} = \vec{E}_1\exp(-i\omega_1t) + \vec{E}_2\exp(-i\omega_2t) + c.c.$ with $\omega_1$ resonant with transition from $n = -1$ to $n = 2$ and $\omega_2$ resonant with transition between $n = 0$ and $n = \pm 1$, where $\vec{E}_1$ has left circular polarization and $\vec{E}_2$ has linear polarization. As a result, the field $\vec{E}_2$ is coupled to both transitions $0 \rightarrow \pm 1$. The partially degenerate 4-wave-mixing interaction generates a right-circularly polarized signal field $\vec{E}_3$ with frequency $\omega_3 = \omega_1 - 2\omega_2$. The signal frequency is in the THz range in a magnetic field of order 1 T. The third-order nonlinear susceptibility corresponding to this process can be calculated using the density matrix approach similarly to \cite{yao}: 
\begin{eqnarray}
\chi^{(3)}(\omega_3) &=& \frac{N\mu_{43}\tilde{\mu}_{41}\tilde{\mu}^*_{32}\tilde{\mu}^*_{21}}{(i\hbar)^3\Gamma_{43}}\left( \frac{\rho_{22}-\rho_{33}}{\Gamma^*_{31}\Gamma^*_{32}} + \frac{\rho_{22}-\rho_{11}}{\Gamma^*_{31}\Gamma^*_{21}}\right. \nonumber \\
 &-&  \left.\frac{\rho_{11}-\rho_{44}}{\Gamma_{42}\Gamma_{41}}+\frac{\rho_{22}-\rho_{11}}{\Gamma_{42}\Gamma^*_{21}}\right).
\label{chi3}
\end{eqnarray}
Here $\tilde{\vec{\mu}}_{mn}$ is defined as $\frac{ie\upsilon_F}{\omega}\langle m|\vec{\sigma}|n \rangle$, which coincides with dipole moment $\vec{\mu}_{mn}$ at resonance; $\Gamma_{mn}$ is the complex dephasing factor between surface LLs $m$ and $n$ \cite{yao}. At resonance, the dephasing factors become real numbers, and we further assume all the detuning rates are the same $\Gamma_{ij}\sim\gamma = 10^{12}$ s$^{-1}$ in the plot below. When all fields are below saturation intensity and the Fermi level is between states 0 and -1, the equivalent 2D third order nonlinear susceptibility for the thin film is $10^{-5}(1/B(T))$ esu. When incident fields increase in intensity, the population differences on the transitions  coupled by the pump fields decrease. As a result, the third order susceptibility drops after the 4 level system gets saturated and the signal intensity decays as well. 

The electric field of the generated signal can be calculated by solving the density matrix equations together with Maxwell's equations. Neglecting the depletion of the pump fields, the relation between the signal field and the nonlinear optical polarization is
\begin{equation} \label{maxw}
\frac{\partial \vec{E}}{\partial z} = i \cdot \frac{2 \pi \omega}{c} \cdot \vec{P}.
\end{equation}
The resulting signal field is then given by $$E_3 = \frac{2\pi i\omega_3}{c}\chi^{(3)}E_1(E^*_2)^2.$$
The  intensity of the nonlinear signal is plotted in Fig.~6(a) as a function of the pump intensity. The maximum signal is reached when the pump fields are of the order of their saturation values $\sim 10^4-10^5$ W/cm$^2$ for the scattering rate $\sim 10^{12}-10^{13}$ s$^{-1}$. The maximum signal intensity decreases as $1/\gamma^3$ with increasing scattering rate. It increases roughly as $B^2$ with the magnetic field, if we ignore the magnetic field dependence of the scattering rate.

\begin{figure}[htb]
\centerline{
\includegraphics[width=8cm]{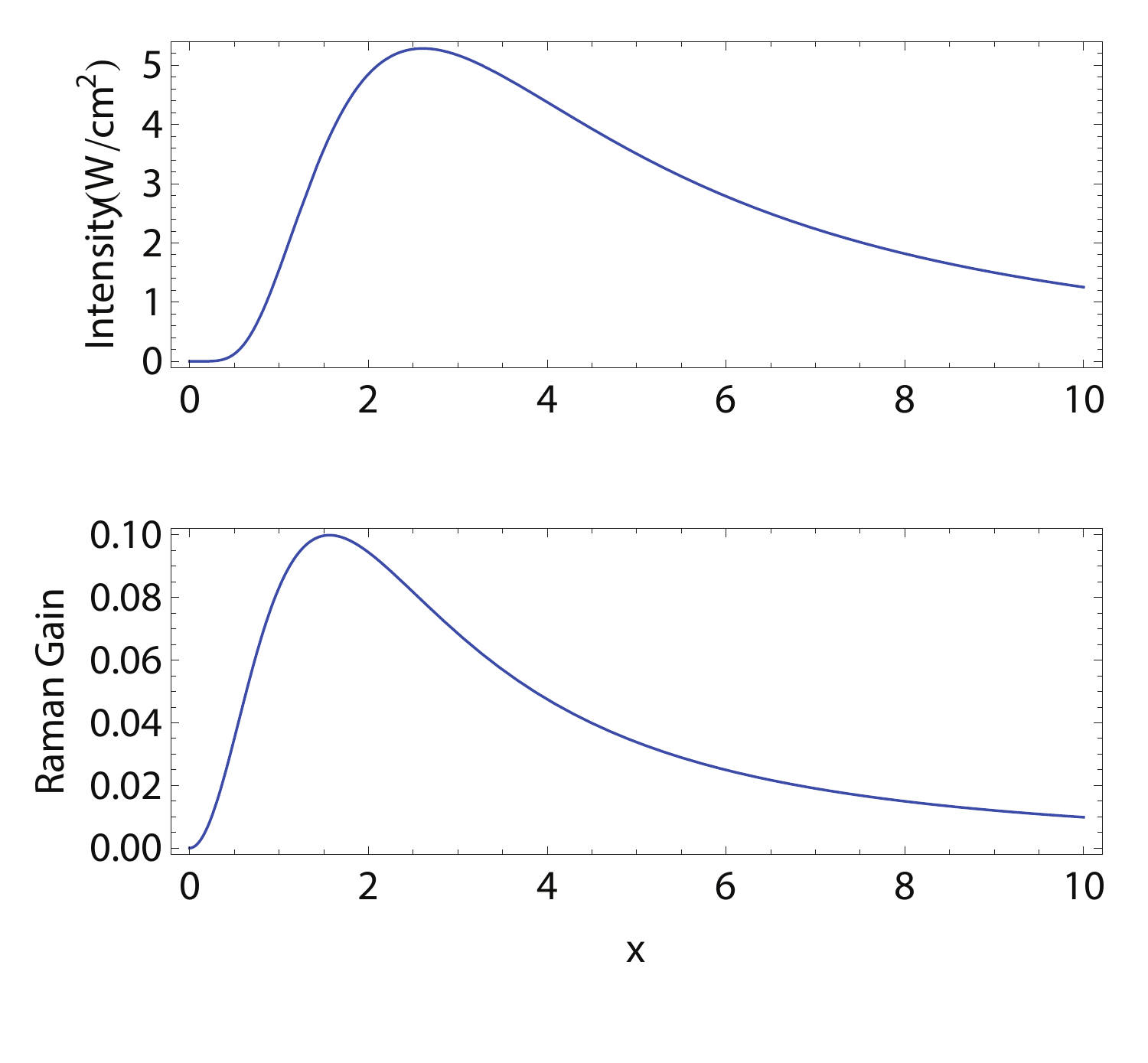}}
\caption{Top panel: Intensity of the four wave mixing signal $E_3$ as a function of the normalized intensity of the pump field $E_1$, $x = I_1/I_{sat}$ where the saturation intensity $I_{sat} \approx 10^4$ W/cm$^2$  in the magnetic field of 1T and for a scattering rate $\gamma = 10^{12}$ s$^{-1}$. The normalized  intensity of the second pump field $E_2$ is taken as $0.6 x$. Bottom panel: the Raman gain $G$ for the field $E_3$ under the same conditions and in the absence of the second field $E_2$.  }
\end{figure}

Another way to generate a coherent THz signal within the level scheme of Fig.~5 is through stimulated Raman Stokes scattering of the pump field $E_1$ into the field $E_3$. In this case one does not need the second pump field $E_2$ and the signal amplification develops exponentially, as  $E_0\exp{G}$, where $G$ is the gain per crossing of the TI film by a normally incident field, which has the form similar to that in graphene  \cite{yao13}:
\begin{eqnarray}
G &=& \frac{2\pi\omega_3 N\mu_{43}\tilde{\mu}_{43}}{\hbar c\Gamma_{43}} \left(n_{43}-\frac{|\Omega_{41}|^2}{\Gamma^*_{31}\Gamma^*_{41}} n_{41}\right)\nonumber\\
&\times & 1/ \left(1 + |\Omega_{41}|^2/(\Gamma_{43}\Gamma^*_{31})\right),
\label{gain1}
\end{eqnarray}
where $\Omega_{41} = \tilde{\mu}_{41} E_1/\hbar$ is the Rabi frequency of the pump field. 
 The gain dependence from the pump intensity is plotted in Fig.~6(b). Similarly to the four-wave mixing case, the gain reaches maximum when the pump field is of the order of the saturation value. 

These results show remarkably high values of the four wave mixing efficiency and Raman gain per monolayer of massless 2D fermions. The nonlinear signal could be enhanced even further by placing the TI film in a  cavity  which increases the effective interaction length of the fields with 2D layers. If the Raman gain is greater than the round-trip losses in a high-Q cavity, one could realize a Raman laser. 

\section{Conclusion}

We have shown that high-quality thin films of topological insulators can be used as basic building blocks for the polarization optics. A nearly complete circular polarization of the incoming radiation can be achieved for a cyclotron resonance with a quality factor $\omega/\gamma \sim 100$. The film thickness can be as small as several nm, and is bounded from below by electron tunneling between the two surfaces which opens the gap. The films also exhibit high third-order optical nonlinearity  resulting in a strong four-wave mixing signal and stimulated Raman scattering in the mid/far-infrared range.  

This work has been supported by NSF Grants OISE-0968405 and EEC-0540832.


\begin{thebibliography}{99}
\bibitem{review} M.Z.Hasan and C.L.Kane, Rev. Mod. Phys. 82, 3045-3067 (2010)
\bibitem{zhang} Chao-Xing Liu, Xiao-Liang Qi, HaiJun Zhang, Xi Dai, Zhong Fang, and Shou-Cheng Zhang, Phys. Rev. B 82, 045122 (2010).
\bibitem{THz} Xiao Zhang, Jing Wang, and Shou-Cheng Zhang, Phys. Rev. B 82, 245107 (2010). 

\bibitem{LLs1} A. A. Schafgans, K. W. Post, A. A. Taskin, Yoichi Ando, Xiao-Liang Qi, B. C. Chapler, and D. N. Basov, Phys. Rev. B 85, 195440 (2012)
\bibitem{LLs2} P. Cheng et al., Phys. Rev. Lett. 105, 076801 (2010).
\bibitem{LLs3} Y. Jiang, Y. Wang, M. Chen, Z. Li, C. Song, K. He, L. Wang, X. Chen,
X. Ma, and Qi-Kun Xue, Phys. Rev. Lett. 108, 016401 (2012). 

\bibitem{aguilar} R. Valdes Aguilar, A.V. Stier, W. Liu, L. S. Bilbro, D. K. George, N. Bansal, L. Wu,
 J. Cerne, A. G. Markelz, S. Oh, and N. P. Armitage,  Phys. Rev. Lett. 108, 087403 (2012). 

\bibitem{hsieh} D. Hsieh et al., Phys. Rev. Lett. 103, 146401 (2009). 

\bibitem{thinfilm_shen} Wen-Yu Shan, Hai-Zhou Lu and Shun-Qing Shen, New Journal of Physics 12, 043048 (2010).
\bibitem{thinfilm_qian} Hai-Zhou Lu, Wen-Yu Shan, Wang Yao, Qian Niu, and Shun-Qing Shen, Phys. Rev. B 81, 115407 (2010).
\bibitem{parameters} H.J.Zhang, C.X.Liu, X.L.Qi, X.Dai, Z.Fang and S.C.Zhang, Nat. Phys. 5, 438 (2009).
\bibitem{LL8} L. D. Landau, E. M. Lifshitz, and L. P. Pitaevski, Electrodynamics of Continuous Media, Butterworth-Heinemann, 2nd edition (1979).
\bibitem{magnetoptics} Wang-Kong Tse and A.H.MacDonald, Phys. Rev. B 82, 161104(R) (2010).
\bibitem{yao} X. Yao and A. Belyanin, Phys. Rev. Lett. 108, 255503 (2012).
\bibitem{tokman} M. Tokman, X. Yao and A. Belyanin, Phys. Rev. Lett. 110, 0774904 (2013).
\bibitem{yao13} X. Yao and A. Belyanin,  J. Phys. Cond. Matt. 25, 054203 (2013). 

\end{thebibliography}
\end{document}